\documentclass[galaxies,article,accept,preprints,oneauthor,dvi2pdf]{Definitions/mdpi} 

\firstpage{1} 
\pubvolume{1}
\issuenum{1}
\articlenumber{0}
\pubyear{2021}
\copyrightyear{2020}
\datereceived{Dec 10, 2021} 
\dateaccepted{Feb 5, 2022} 
\datepublished{Feb 8, 2022} 
\hreflink{https://doi.org/} 

\Title{Plasma Diagnostics in the Era of Integral Field Spectroscopy}

\TitleCitation{Plasma Diagnostics in the Era of IFS}

\Author{Toshiya Ueta$^{1,2,\dagger}$\orcidA{}}

\AuthorNames{Toshiya Ueta}

\AuthorCitation{Ueta, T.}

\address{%
$^{1}$ \quad University of Denver, 2112 E Wesley Ave., Denver, CO 80208, U.S.A.; toshiya.ueta@du.edu\\
$^{2}$ \quad Okayama Observatory/Kyoto University, Honjo, Kamogata, Asakuchi, Okayama, 719-0232, Japan}

\corres{Correspondence: toshiya.ueta@du.edu}

\firstnote{Japan Society of the Promotion of Science Invitational Fellow (FY 2020; long-term)} 

\abstract{%
To understand the physical conditions of various gaseous systems,
plasma diagnostics must be performed properly.
To that end, it is equally important to have 
extinction correction performed properly, 
even before performing plasma diagnostics. 
This means that the physical conditions of the target sources
-- the very quantities to be derived via plasma diagnostics --
must be known even before performing extinction correction,
because the degree of extinction is determined 
by comparing the observed spectra of the target sources
with their theoretically predicted counterparts.
One way to resolve this conundrum is 
to perform both extinction correction and plasma diagnostics 
together by iteratively seeking a converged solution.
In fact, if these analyses are performed self-consistently, 
a converged solution can be found 
based solely on well-calibrated line intensities,
given the adopted extinction law and the $R_V$ value.
However, 
it is still rare to find these analyses 
done numerically rigorously without unnecessary analytical approximations 
from start to finish.
In this contribution for the APN\,8e conference,
we would like to review this convoluted problem
and sort out critical issues 
based on the results of our recent experiments.
It appears 
that the convoluted theoretical and observational progresses 
exacerbated by the highly numerical nature of these analyses
necessitated a number of analytical simplifications
to make the problem analytically tractable in the pre-computer era
and that such analytical simplifications still remain rampant
in the literature today
even after ample computational resources became readily available.
Hence, the community is encouraged to do away with this 
old habit of sidestepping numerical calculations 
that was a necessary evil in the past.
This is especially true in the context of spatially-resolved 
2-D spectroscopy, which obviously conflicts with 
the uniformity assumption often blindly inherited from 1-D spectroscopy.}

\keyword{Astronomical methods; Plasma diagnostics; Extinction Correction} 

\begin{document}

\section{Introduction}

To our understanding of the physical conditions of various astrophysical gaseous systems,
it is fundamental to perform plasma diagnostics
\cite{osterbrock1989,osterbrock2006}.
The relative strengths of various diagnostic emission lines determine 
the excitation states of specific gaseous species, 
yielding their electron density ($n_{\rm e}$) and temperature ($T_{\rm e}$), 
and subsequently, metal abundances \cite{peimbert2017,nicholls2020}.
Meanwhile, it is equally fundamental to perform extinction correction
caused by both the interstellar and circumsource components, 
so that genuinely unattenuated intrinsic spectra are available 
for plasma diagnostics \cite{draine2003,salim2020}.
This is the quintessence of observational astronomy, 
in which all measurements made from a distance are affected by extinction.

However, the determination of extinction, 
$c(\lambda)$\footnote{The extinction at $\lambda$, $c(\lambda)$, is the base-10 power-law index to describe the reduction of the intrinsic flux, $I_0(\lambda)$, to the observed flux, $I(\lambda)$, by $I(\lambda)/I({\rm H}\beta) = I_0(\lambda)/I_0({\rm H}\beta) \times 10^{-\left(c(\lambda)-c({\rm H}\beta)\right)}$. It is customary to scale the arbitrary extinction law, $c(\lambda)$, relative to $c$ at ${\rm H}\beta$.}, 
is not a trivial task.
In plasma diagnostics that is presently widely practiced,
the amount of extinction is usually determined 
by comparing observed diagnostic H\,{\sc i} recombination line ratios 
(e.g.\ $I({\rm H}\alpha)/I({\rm H}\beta)$ and $I({\rm H}\beta)/I({\rm H}\gamma)$)
with the corresponding theoretical counterparts 
(i.e.\ unattenuated intrinsic line ratios)
\cite{osterbrock1989,osterbrock2006}. 
Such unattenuated line ratios are simply dependent on 
the specific $n_{\rm e}$ and $T_{\rm e}$ of line-emitting gas 
in the target sources and can be computed {\em numerically} 
given the desired complexity of the atomic physics
\cite{hummer+1987,storey+1995}.
Needless to say, $n_{\rm e}$ and $T_{\rm e}$ are the very quantities 
to be determined via plasma diagnostics 
with {\em extinction-corrected} line ratios. 
Hence, this conundrum is a classic catch-22, or, chicken-and-egg situation.

In the mean time, in typical plasma diagnostics performed today, 
atoms are usually represented as $n$-level energy states.
Thus, $n_{\rm e}$ and $T_{\rm e}$ are determined 
from a set of equilibrium equations for the adopted $n$-level system. 
In these equilibrium equations, 
the collisional excitation coefficient has the
$n_{\rm e} \sqrt{T_{\rm e}} \exp(-\Delta E/kT_{\rm e})$ dependence 
(where $\Delta E$ is the energy difference between any two levels) 
and the collisional de-excitation coefficient has 
the $n_{\rm e} \sqrt{T_{\rm e}}$ dependence,
while the radiation de-excitation coefficient has no dependence on 
$n_{\rm e}$ and $T_{\rm e}$ {\em to the first order} 
\cite{pradhan+2015}.
Because these equations with exponential functions are transcendental, solutions are obtained only {\em numerically}.

Therefore, to overcome the conundrum pointed out above, 
both plasma diagnostics and extinction correction 
ought to be performed together {\em as a streamlined iterative numerical process}.
This is, in fact, suggested even when the present procedure of 
plasma diagnostics was introduced to the community
\cite{osterbrock1989,osterbrock2006}.
However, what has been traditionally exercised in the literature 
is to adopt a number of approximations 
to make the problem analytically tractable 
(which was done originally {\em for instructional purposes} \cite{osterbrock1989,osterbrock2006}). 
Moreover, {\em ad hoc} $n_{\rm e}$ and $T_{\rm e}$ are often adopted 
to force a value of $c({\rm H}\beta)$,
or even an {\em ad hoc} $c({\rm H}\beta)$ value may be adopted
to skip extinction correction altogether.

In recent years, spatially-resolved 2-D plasma diagnostics have 
been becoming very relevant in many branches of astronomy,
especially with the increasing availability of integral field 
spectroscopy (IFS; e.g.\ \cite{walsh+2020}).
For extended targets, 
$n_{\rm e}$ and $T_{\rm e}$ (and hence, $c({\rm H}\beta)$)
are of course expected to vary spatially. 
However, even for such spatially extended targets, 
the spatial variation of $n_{\rm e}$ and $T_{\rm e}$ 
(and $c({\rm H}\beta)$) does not seem to have been 
considered carefully enough. 
Often, uniform $n_{\rm e}$ and $T_{\rm e}$ (or $c({\rm H}\beta)$)
are adopted across extended target sources,
instead of performing calculations 
at each detector element (i.e.\ pixel or spaxel),
defeating the purpose of spatially-resolved IFS observations.
In such cases, self-consistency is regrettably nil.

The subtlety of self-consistency 
between extinction correction and plasma diagnostics 
seems to have been lost in translation, 
most likely because 
these analyses are often regarded as two separate problems. 
Hence, self-consistency between these two sets of 
$n_{\rm e}$ and $T_{\rm e}$ in extinction correction and plasma diagnostics 
is rarely scrutinized, let alone guaranteed.
Consequently, such inconsistencies would usually invite uncertainties, 
albeit inadvertently.

\section{Typical Procedure in the Literature}

Because this problem turns out to be rather convoluted,
let us first sort out critical points by closely examining 
the procedure of extinction correction and plasma diagnostics 
typically employed in the literature.
Extinction correction begins with adopting an extinction law
and the associated $R_V$ value to scale the extinction law.
Then, the extinction $c$ at a reference wavelength 
(customary at H$\beta$, i.e.\ $c(\beta)$) is determined 
by comparing observed diagnostic H\,{\sc i} recombination line 
ratios (most commonly $I({\rm H}\alpha)/I({\rm H}\beta)$
and/or $I({\rm H}\gamma)/I({\rm H}\beta)$) with the
theoretical predictions (i.e.\ unattenuated line ratios).
Here, the theoretical H\,{\sc i} recombination line ratios 
are nothing but functions of $n_{\rm e}$ and $T_{\rm e}$ 
\cite{hummer+1987,storey+1995}.
Hence, to guarantee self-consistency between extinction correction 
and plasma diagnostics,
the input $n_{\rm e}$ and $T_{\rm e}$ that define 
the unattenuated line ratios for comparison in extinction correction
must be consistent with the resulting $n_{\rm e}$ and $T_{\rm e}$
to be derived via plasma diagnostics.

In the literature, 
there is often a reference to the "canonical" 
$I_0({\rm H}\alpha)/I_0({\rm H}\beta)$ ratio at this point in the process.
The most often quoted 
ratio is probably 2.858, which is true only when 
$n_{\rm e} = 10^3$\,cm$^{-3}$ and $T_{\rm e} = 10^4$\,K
\cite{hummer+1987,storey+1995}.
This is totally misleading for uninitiated.
The $I_0({\rm H}\alpha)/I_0({\rm H}\beta)$ ratio is simply 
a function of $n_{\rm e}$ and $T_{\rm e}$,
and there is {\em no such thing as the canonical ratio}.
Because no specific $(n_{\rm e}, T_{\rm e})$ values would warrant 
any canonicality for the resulting $I_0({\rm H}\alpha)/I_0({\rm H}\beta)$ ratio,
the ratio simply has to be computed 
based on the given $n_{\rm e}$ and $T_{\rm e}$.
It appears that this unwarranted canonicality of the $I_0({\rm H}\alpha)/I_0({\rm H}\beta)$ ratio 
often referenced in the literature introduced an unfortunate disconnect between 
extinction correction and plasma diagnostics,
because inexperienced tend to blindly quote 
the "canonical" $I_0({\rm H}\alpha)/I_0({\rm H}\beta)$ ratio
and be done with it
rather than fully appreciating its $(n_{\rm e}, T_{\rm e})$ dependence.

This is obviously a bad start for the subsequent
plasma diagnostics, which require extinction-corrected
line strengths as inputs.
If $c({\rm H}\beta)$ is not computed according to the correct 
$I_0({\rm H}\alpha)/I_0({\rm H}\beta)$ ratio via proper $n_{\rm e}$ and $T_{\rm e}$,
the resulting extinction-corrected spectrum is already compromised.
The uncertainty caused by this incorrect $c({\rm H}\beta)$ would not just scale
with it because $c(\lambda)$ varies with the wavelength.
Also, 
the uncertainty in $c(\lambda)$ would amplify the uncertainty in the resulting line strengths by $\ln(10)$.
because $c(\lambda)$ is the base-10 power index.
Hence, any line ratios measured from such an erroneously extinction-corrected spectrum are obviously faulty,
and the results of plasma diagnostics undermined by such erroneous inputs would be clearly unreliable.

There are additional sources of inconsistency in plasma diagnostics.
The high point of plasma diagnostics is determining 
$n_{\rm e}$ and $T_{\rm e}$
by pinpointing where two diagnostic curves 
of the measured line ratios intersect 
in the $n_{\rm e}$-$T_{\rm e}$ plane.
In general, a line ratio can be computed 
as a function of $n_{\rm e}$-$T_{\rm e}$
based on the equilibrium equations of the adopted $n$-level system
for the atomic species in question.
As the equilibrium equations are transcendental in $n_{\rm e}$ and $T_{\rm e}$, 
the determination of the intersection between two 
diagnostic line ratio curves has to be done numerically.

In practice, it is conventional to use the so-called 
$n_{\rm e}$- and $T_{\rm e}$-diagnostic line ratios
as a pair.
On the one hand, 
the $T_{\rm e}$-diagnostic line ratios are those having only 
weak $n_{\rm e}$ dependence.
If one pushes the low-density limit (i.e. taking $n_{\rm e} \rightarrow 0$),
the line ratio can be expressed analytically as a function of $T_{\rm e}$ only.
On the other hand, 
the $n_{\rm e}$-diagnostic line ratios are those having only 
weak $T_{\rm e}$ dependence.
The line ratio varies with $n_{\rm e}$ between two asymptotic values
only within a specific range of $n_{\rm e}$.
However, where this range of $n_{\rm e}$ falls is weakly dependent on $T_{\rm e}$.
This step-function behavior of the ratio with $n_{\rm e}$ is often shown 
by a plot for a specific $T_{\rm e}$ case {\em for instructional purposes} 
(e.g.\ \cite{osterbrock1989,osterbrock2006}).
Then, such a plot, and especially an analytic translation of it, can become prevalent 
in the literature with the original $T_{\rm e}$ specificity forgotten.
Hence, these weak $n_{\rm e}$- and $T_{\rm e}$-dependences 
in the corresponding $T_{\rm e}$- and $n_{\rm e}$-diagnostics
wither away 
as the ease of use of such analytical expressions is favored over 
the cumbersomeness of rigorous numerical calculations.

In addition, the choice of $n_{\rm e}$- and $T_{\rm e}$-diagnostic line ratio pairs
can be a source of inconsistency. 
Ideally, emission lines adopted as a diagnostic pair should originate from 
the same region of the target object along the line of sight
so that plasma diagnostics do actually probe $n_{\rm e}$ and $T_{\rm e}$ 
of this region.
This practically means that these line emission should be of roughly 
the same transition energies
(e.g.\ [N\,{\sc ii}], [S\,{\sc ii}], and [O\,{\sc ii}] lines for low-excitation regions,
and [O\,{\sc iii}], [Cl\,{\sc iii}], and [Ar\,{\sc iv}] lines for high-excitation regions;
\cite{osterbrock1989,osterbrock2006}).
In the literature, however, diagnoatic line pairs do not seem to be selected 
as deliberately as they ought to be.
The resulting $n_{\rm e}$ and $T_{\rm e}$ values, therefore, often seem 
to be the mere average of as many permutations of diagnostic line pairs 
as possible.
If lines associated with very different transition energy regimes are used 
together as a diagnostic line pair, 
the resulting $n_{\rm e}$ and $T_{\rm e}$ 
{\em may not represent any part of the target} along the line of sight.

\section{PPAP: Proper Plasma Analysis Practice}

On the whole,
the discussion above can be distilled into the following four major 
points of consideration when extinction correction and plasma diagnostics 
are to be performed effectively as a single integrated procedure: 
\begin{enumerate}
\item keep track of $n_{\rm e}$ and $T_{\rm e}$ from start to finish 
in order not to get distracted by secondary derivatives such as the 
diagnostic H\,{\sc i} recombination line ratios 
that are simply functions of $n_{\rm e}$ and $T_{\rm e}$;

\item stick to rigorous numerical calculations without
resorting to analytical approximations that may
be true only for specific circumstances;

\item take into account the physical conditions of the regions of the 
target source to select appropriate diagnostic line pairs 
that actually represent the regions to be probed; and

\item execute all calculations at each detector element 
to fully account for spatial variation when the target object is extended.
\end{enumerate}

While performing an exhaustive investigation in the literature is practically not possible, 
it appears to be rare to find previous works of extinction correction and plasma diagnostics 
in which all four of the above are rigorously implemented.
In the literature, it is often {\em ambiguous} as to how analyses were done exactly,
mainly because in works of plasma diagnostics extinction correction is typically mentioned only in passing.
To that end, we have recently carried out a small proof-of-concept experiment, 
in which all of the above are carefully carried out in performing 
extinction correction and plasma diagnostics 
(dubbed proper plasma analysis practice (PPAP), \cite{ueta+2021}).

The experiment was done with a set of HST/WFC narrowband images
of the NW quadrant of the PN NGC\,6720, obtained from the data archive \cite{odell+2013}.
This experiment was performed as a demonstration follow-up of another work,
in which a new algorithm was developed to isolate multiple emission line maps 
from a set of narrowband images whose filter profiles overlap with each other
\cite{ueta+2019}.
In particular for the NGC\,6720 data set,
the H$\alpha$ 6563\,{\AA} and [N\,{\sc ii}] 6548/83\,{\AA} maps
were isolated from the F656N and F658N images,
while 
the [S\,{\sc ii}] 6717/31\,{\AA} maps were recovered from the F673N, FQ672N, and FQ674N images
\cite{ueta+2019,ueta+2021}.

\begin{figure}[hb]
\centering
\includegraphics[width=0.6\textwidth]{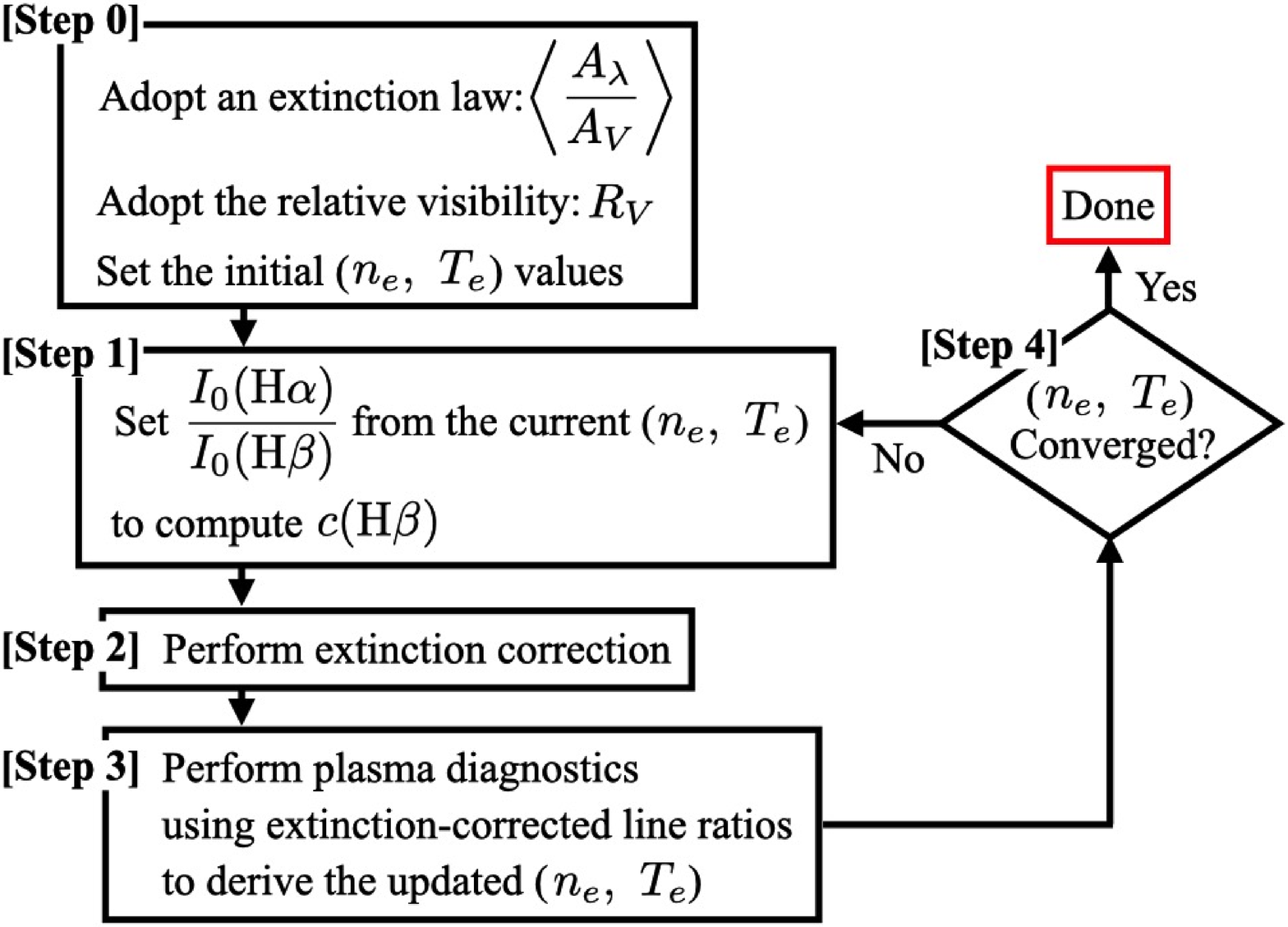}
\caption{A flow chart of PPAP \cite{ueta+2021}, through which converged 
$n_{\rm e}$ and $T_{\rm e}$ are sought by a streamlined numerical iterative process
given the selection of the extinction law and the corresponding 
$R_{V}$ value as well as the diagnostic line pairs.\label{fig:ppap}}
\end{figure}   

As presented in a flow chart (Fig.\,\ref{fig:ppap}), PPAP is a honest no-nonsense 
implementation of extinction correction and plasma diagnostics
aiming at doing away with analytical approximations 
adopted previous to the modern computer era.
After selecting an extinction law and the associated $R_V$ value toward the target object,
the rest is essentially an autopilot of numerical evaluations 
that seeks a converged solution of $(n_{\rm e}, T_{\rm e})$
between extinction correction 
(dictated by the present values of $n_{\rm e}$ and $T_{\rm e}$
in evaluating the theoretical predictions of the diagnostic H\,{\sc i} recombination line ratios;
e.g.\ \cite{storey+1995})
and plasma diagnostics 
(based on the observed ratios of diagnostic lines whose transition energies 
are appropriate for the regions to be probed in the target objects).

An abridged list of important findings in this PPAP experiment includes;
\begin{enumerate}
\item the $n_{\rm e}$ and $T_{\rm e}$ distributions are not at all uniform, 
and so are the derived 
diagnostic H\,{\sc i} recombination line ratio (e.g.\ $I_0({\rm H}\alpha)/I_0({\rm H}\beta)$) and 
$c({\rm H}\beta)$ distributions;

\item if a constant $c({\rm H}\beta)$, $I_0({\rm H}\alpha)/I_0({\rm H}\beta)$, or $(n_{\rm e}, T_{\rm e})$ were assumed in extinction correction, 
spatially-varying over- and under-correction of extinction would have occurred
as the degree of attenuation could have been off by several tens of \% in the observed part of the nebula,
compromising the resulting "extinction-corrected" emission line maps;

\item the dust distributions ($= c({\rm H}\beta)$ distributions)  can be obtained solely 
from the optical spectral images
(i.e.\ no need to obtain separate thermal dust emission maps in the infrared); 

\item the relative ionic abundance distributions of 
$n({\rm N}^+)/n({\rm H}^+)$ separately derived from 
each of the two diagnostic lines
in the [N\,{\sc ii}] 6583\,{\AA} and [N\,{\sc ii}] 5755\,{\AA} 
turned out to be identical within uncertainties,
and so did the $n({\rm S}^+)/n({\rm H}^+)$ distributions
derived from the [S\,{\sc ii}] 6717\,{\AA} and [S\,{\sc ii}] 6731\,{\AA} line maps; and

\item simulated analyses tolerating the uniform $(n_{\rm e}, T_{\rm e})$ (or $c({\rm H}\beta)$) 
distribution in extinction correction would result in spatially varying uncertainties 
at several tens of \% in the derived $(n_{\rm e}, T_{\rm e})$ and relative ionic abundances distributions.

\end{enumerate}

The significant take-away from this experiment is that 
results of plasma diagnostics could be off by several tens of \%
if PPAP is not strictly followed.
Again, PPAP is a straightforward implementation of extinction correction and plasma diagnostics with no frills
as suggested from ages ago by many (e.g.\,\cite{osterbrock1989,osterbrock2006,peimbert2017,nicholls2020}).
As shown in the flow chart (Fig.\,\ref{fig:ppap}), 
it is as simple as {\em performing both extinction correction and plasma diagnostics completely numerically
as a streamlined iterative process to seek the converged solution of $(n_{\rm e}, T_{\rm e})$
 without resorting to any analytical approximations and generalizing assumptions}.
There is neither new theory nor new numerical procedure to adopt.
The only thing necessary is an honest implementation of the existing analyses of 
extinction correction and plasma diagnostics at face value.

It is not too difficult to imagine that seeking $(n_{\rm e}, T_{\rm e}$) completely numerically for convergence 
used to be too cumbersome to perform in the past, especially when computational resources were scarce.
Hence, it is understandable that 
a number of analytical approximations had to be adopted as a necessary evil in the past
to make the whole procedure analytically tractable. 
However, such temporary measures of the pre-computer era
are still regularly practiced even today when sufficient computational resources are readily available.
Therefore, there is only our negligence to blame.
It is time to do away with this old habit of sidestepping numerical calculations,
simply because we can now perform all these numerical calculations at ease.

 \section{Historical Perspective}

Here, to mend our own negligence,
let us briefly explore the historical developments around extinction correction and plasma diagnostics in the literature
and gain more insights as to why 
\begin{enumerate}
\item extinction correction has not been incorporated as closely as it should have been in plasma diagnostics, and

\item the community has not yet managed to have gone fully numerical,
\end{enumerate}
which are the very questions that concern with the main theme of PPAP.

According to Aller \cite{aller1999}, 
Menzel and his collaborators performed the pioneering work to establish 
the process of plasma diagnostics based on spectral line intensities 
(which is presently known as the "Direct Method")
via a series of 18 papers from the 1930s to 1940s
(e.g.\,\cite{menzel1937,baker+1938,aller+1945}).
It was still when observations were made by 
"eye estimates" from photographic plates and when the atomic parameters were largely unknown.
Hence, Aller himself stated even in 1951 that 
"Because of the uncertainties in the collisional cross-sections, 
we are unable to derive ionic abundances and electron temperatures from the nebular line intensities, 
nor does it seem worth while to calculate electron densities, since the nebular surface brightnesses
and distances are so poorly known" \cite{aller1951}.
It was only some 70 years ago.

Despite such adversities, Aller and collaborators pressed on in the 1950s and 1960s
as modern techniques gradually improved observational uncertainties \cite{aller1954}.
It was around that time when the need for extinction correction for spectral lines was pointed out \cite{aller+1955}.
Burgess and Seaton were the early adopters of extinction correction
in the context of PN plasma diagnostics \cite{burgess1958,seaton1960},
following the ISM extinction work by Whitford \cite{whitford1958}.
Unfortunately, however, 
{\em discrepancies between observations and theoretical predictions were larger 
than extinction alone could account for},
because the recombination theory at the time did not take into account the collisional effects.

Then, it took two more decades through the 1970s and 1980s
until the collisional effects in the recombination theory were fully taken into consideration,
first implemented by Brocklehurst \cite{brocklehurst1971} for specific cases and
later generalized by Hummer and Storey \cite{hummer+1987}.
This theoretical development took nearly two decades not only because of 
the technical difficulty but also because 
{\em observational uncertainties at the time were still often too large to corroborate theoretical predictions}.
Nevertheless, 
with both of the $n_{\rm e}$ and $T_{\rm e}$ dependences involved in plasma diagnostics established, 
it was the end of the 1980s when the {\em de facto} standard textbook for the subject matter 
was authored by Osterbrock \cite{osterbrock1989}, 
which many existing methods of plasma diagnostics, including PPAP, are based on.

Concerning PPAP, 
the works by Hummer and Storey \cite{hummer+1987} and Storey and Hummer \cite{storey+1995},
for example, essentially established 
a way to connect extinction correction and plasma diagnostics seamlessly
via the Balmer line ratios,
because these ratios turned out to be easier to establish observationally 
than the Balmer decrement and Paschen-to-Balmer ratios that were typically 
used before then \cite{miller+1972}.
However, as mentioned above, instead of thoroughly exploring the $n_{\rm e}$ and $T_{\rm e}$ dependences
in extinction correction and plasma diagnostics numerically \cite{hummer+1987,storey+1995},
the "canonical" Balmer line ratio was introduced to sidestep extinction correction
even though there was nothing to vouch for the claimed canonicality.
This happened most likely because 
the advantage of sidestepping the volume of numerical computation necessary 
in following the $n_{\rm e}$ and $T_{\rm e}$ dependences of extinction correction and plasma diagnostics 
rigorously
outweighed 
the disadvantage of not doing so 
given 
the relative shallowness of the $n_{\rm e}$ and $T_{\rm e}$ dependences on the Balmer line ratio
\cite{hummer+1987,storey+1995}
and 
the computational resources typically available at the time.
It is true that computational resources were still commodities in the 1980s and 
even in the 1990s.

While the consideration just above may answer the first of the two questions raised at the beginning of this section,
the second question is very puzzling:
why has the community not yet gone completely numerical on this matter two decades later?
The PPAP experiment was done using a laptop \cite{ueta+2021}.
There exist many codes of plasma diagnostics both proprietary and in the public domain,
including {\sc NEAT} \cite{wesson+2012}, which can even propagate uncertainties 
from line flux measurements to the derived abundances 
and {\sc PyNeb} \cite{pyneb}, a Python implementation of the latest {\sc nebular} lineage of the IRAF fame,
which popularized the diagnostics \cite{shaw+1995}.
Hence, the availability of computational resources cannot be an issue.

Looking back on the historical developments briefly summarized above, 
there seems to be a recurring pattern of
{\em competition between observational uncertainties
and the cost of the adopted mode of analysis.} 
At the very beginning in the 1930s and 1940s, 
observational uncertainties at the time made it look as if
plasma diagnostics was impossible. 
During the 1950s and 1960s, 
only the $T_{\rm e}$ dependence was considered 
because any consideration of the $n_{\rm e}$ dependence,
even including extinction correction,
was buried under observational uncertainties.
Through the 1970s and 1980s, 
full consideration of the $n_{\rm e}$ and $T_{\rm e}$ dependence 
was yet again stagnated by observational uncertainties.
This repeating pattern might have affected the collecitve psyche of the community
to shy away from going fully numerical. 
As we just saw, 
the "canonical" Balmer line ratio gained popularity in the following decades
because uncertainties caused by not following the $n_{\rm e}$ and $T_{\rm e}$ 
dependences rigorously were deemed tolerable.
For some unkown reason, the community seems to have always assumed,
and have been assuming,  
that uncertainties that result by not going fully numerical would still be minor.
This may well be the collecitve psyche of the community
influenced by the constant need to assess the balance 
between observational uncertainties
and the cost of the adopted mode of analysis for the past 70 years or so.

\section{Plasma Diagnostics in the Era of IFS}

It is possible to keep speculating as to why a fully numerical approach 
such as PPAP has rarely been attempted 
to simultaneously seek a self-consistent converged solution for both extinction correction and plasma diagnostics iteratively.
However, it does not seem to be figured out ever behind the rich but convoluted history
of extinction correction and plasma diagnostics. 
Plus, there does not seem to be much point in doing so: 
we certainly did not review the history to point the finger at anyone.
PPAP is simply one genuinely sensible adaptation of extinction correction and plasma diagnostics,
aiming at performing these analyses with the least amount of approximations and assumptions. 
As a result, what PPAP requires is just the input spectral imaging data set 
plus a choice of the extinction law and the associated total-to-selective extinction, $R_V$, 
both of which can nowadays be set with a reasonable amount of confidence for any given target source.

Our predecessors simply had to be creative in dealing with these analyses 
that actually require fully numerical approaches 
when they did not possess appropriate computational resources.
Now that each one in the community has a decent amount of computational resources,
it is time to abolish all of such approximations and assumptions that may have been needed in the past but not any longer.
This is simply because we can do so and because we can get less uncertain results by doing so.
Hence, there really does not seem any reason not to do so.
In fact, it must be done if target sources exhibit spatial variations at 10\,\% or less
because the present "canonical" procedure ladened with approximations is prone to uncertainties at tens of \%.
Therefore, the community is encouraged to {\em do away with this old habit} that would do more harm than good 
and take up on PPAP or alike especially in the context of extinction correction and plasma diagnostics by means of spatially-resolved 2-D integral field spectroscopy.



\vspace{6pt} 




\funding{This research was partially supported by the Japan Society for the Promotion of Science (JSPS) through its invitation fellowship program (FY2020, long-term) awarded to T.U.}

\appendixtitles{no} 

\end{paracol}
\reftitle{References}




\end{document}